\documentclass{PoS}
\usepackage{amsmath,upgreek}

\PoS{PoS(LAT2005)019}

\title{Performance of machines for lattice QCD simulations}

\ShortTitle{Performance of machines for lattice QCD simulations}

\author{\speaker{Tilo Wettig}\\ \\
  Institute for Theoretical Physics\\
  University of Regensburg\\
  93040 Regensburg, Germany\\ \\
  E-mail: \email{tilo.wettig@physik.uni-regensburg.de}}

\abstract{We review the architecture of massively parallel machines
  used for lattice QCD simulations and present benchmarks for the
  performance of popular algorithms on these platforms.  We cover
  commercial supercomputers, PC clusters, and custom-designed
  machines.  We also speculate on future developments.}

\FullConference{XXIIIrd International Symposium on Lattice Field Theory\\
  25-30 July 2005\\
  Trinity College, Dublin, Ireland}

\begin{document}

\section{Introduction}
\label{sec:intro}

Lattice gauge theory is a rather mature field of research.  There are
many well-defined questions, of both fundamental and phenomenological
importance, to which we can obtain definite answers provided
sufficient computing power is available.  While steady progress has
been made on the theoretical and algorithmic fronts, advances in
computing technology have probably made the largest contribution to
the performance increase of lattice QCD simulations in the past.  This
is likely to remain the case in the future.

Clearly, the computational demands of large-scale lattice QCD
simulations \cite{kennedy} can only be satisfied by massively parallel
machines.  There are many considerations contributing to the design
(or purchase) and the operation of such a machine, most of which will
be discussed in this review.  From a funding agency's point of view,
leaving aside political and other nontechnical issues, the bottom line
is the cost-effectiveness of the machine, i.e., the ``science output''
per unit of currency.  All other details being equal, such as the
particular physical quantity to be computed and the algorithm
employed, a good measure for this is the sustained performance (to be
defined below) of a parallel machine per unit of currency.  Of course,
things are not as simple as a single number, and one of the purposes
of this paper is to elucidate some of the important details that are
hidden behind such a number.

The structure of this paper is as follows.  In Sec.~\ref{sec:basics},
we introduce some basic aspects of parallel computing to make the
presentation accessible to the non-expert reader.  We then summarize
performance-relevant issues that play a role in the design of parallel
machines in Sec.~\ref{sec:design}.  This is followed by the main part
of the paper, Sec.~\ref{sec:machines}, in which we introduce popular
machine architectures used by the lattice community (commercial
supercomputers, PC clusters, and custom-designed machines) and present
benchmarks for the performance of lattice QCD algorithms on these
platforms.  Section~\ref{sec:future} speculates on future
developments, and conclusions are drawn in Sec.~\ref{sec:concl}.

Lattice QCD practitioners are not only expert users of massively
parallel machines, they are also actively involved in the design and
development of new parallel computing technology, both in the hardware
and in the software sector.  There is enormous talent, expertise, and
experience available in our community (which, unfortunately, is
typically operating under rather tight financial constraints).  This
review is in large part a reflection of the hard and ingenious work of
these researchers.

\section{Parallel computing basics}
\label{sec:basics}

In general, parallelization of a computational task means the
distribution of various parts of this task onto several resources.
Parallelism can be exploited at several levels.  For example, modern
CPUs utilize the so-called instruction-level parallelism of a program
by pipelining the various stages of an instruction and executing them
in parallel.
In a more restricted sense, we often use the term parallelization to
mean the numerical solution of a problem on several processors, with
each processor working on a subvolume of the problem.  These
processors need to communicate through some sort of interconnection
network.  Lattice QCD is particularly amenable to this kind of
parallelization since (a) one typically uses a regular hypercubic grid
that can easily be split into identical subvolumes, (b) the boundary
conditions are simple, and (c) the communication patterns between
compute nodes are uniform and predictable.  For example, a
state-of-the-art simulation with dynamical fermions on a
$32^3\times64$ global volume could be done on a parallel machine with
8,192 nodes and a local (i.e., per-node) volume of $4^4$.  Since the
same program is executed on all nodes, the machine is said to operate
in SPMD (single program, multiple data) mode.

The performance of a parallel machine depends on the performance of
its hardware components (processor, memory susbsytem, and
communications network) and on the software that runs on the machine
(application code, libraries, operating system).  The overall
performance can be quantified in terms of peak and sustained
performance.  The peak performance is a single number that is simply
the number of processors (which we assume to be identical) times the
peak performance of a single processor.  The latter is the product of
the clock frequency and the number of double-precision floating point
operations (Flops) that can be executed in a single clock cycle.  For
example, a cluster of 256 Pentium-4 PCs (two Flops per cycle) running
at 3 GHz has a peak performance of 1.5 TFlop/s.  What really matters,
of course, is the sustained performance, which is the actual number of
Flops executed (on average) by the application code per clock cycle,
again multiplied by the clock frequency and the number of processors.
The sustained performance is often quoted as percentage of peak, which
gives an indication of how well a particular application performs on a
given machine.  However, as mentioned above, it is the sustained
performance in Flop/s per unit of currency that is the best measure of
the cost-effectiveness of a machine.

We can now think of defining a ``cost-effectiveness function'' that
depends on several control parameters which  we can tune to maximize
this function.  Examples of such control parameters are\\[-6mm]
\begin{itemize}\itemsep-1mm
\item clock frequency,
\item number of Flops per clock cycle (e.g., SIMD vector operations),
\item number of processors,
\item percentage of peak sustainable,
\item expenses for power, cooling and floor space.\\[-6mm]
\end{itemize}
Of course, there are dependencies between these parameters, e.g., a
higher frequency in general means higher power consumption, and a
larger number of processors often means smaller sustained performance
because more communication is needed.  The most interesting and
nontrivial parameter is clearly the sustained performance on which we
will mainly focus in the following.  It
depends on a variety of factors such as\\[-6mm]
\begin{itemize}\itemsep-1mm
\item stalls due to memory access (if the data are not in cache),
\item stalls due to communication between processors,
\item software overhead, e.g., for operating system or communication
  calls,
\item imbalance of multiply and add in the algorithm (e.g., some CPUs
  such as PowerPC and Itanium provide so-called fused multiply-add, or
  FMA, instructions, and if the numbers of multiplies and adds differ,
  the peak performance cannot even be obtained in theory).\\[-6mm]
\end{itemize}
Clearly, hardware and software should be designed or selected to
minimize the dead time and to maximize the sustained performance.
There are many details that need to be taken into account, but the
final result can be exhibited in a rather simple way in a so-called
scalability plot, where the sustained performance is plotted as a
function of the number of processors participating in the calculation.
Such a plot contains the essential information about the quality of a
machine.  However, two kinds of scalability plots need to be
distinguished:\\[-6mm] 
\begin{itemize}\itemsep-1mm
\item ``weak scaling'': Here the local volume is kept fixed, and thus
  the global volume, i.e., the problem size, increases linearly with
  the number of processors.
\item ``strong scaling'': Here the global volume is kept fixed so that
  the local volume decreases with the number of processors.\\[-6mm]
\end{itemize}
The latter case is the more interesting one since we typically want to
solve a fixed physical problem in the shortest possible wall clock
time.  (To give a simple example from another field, the weather
forecast for tomorrow better be done by today.)  Ideally, one would
like to obtain linear scaling, i.e., a linear increase of the
sustained performance with the number of processors.  However, a
decreasing local volume results in two competing effects: (1) more (or
even all) of the data associated with the local subvolume will fit in
cache (or on-chip memory), which is good, and (2) the
surface-to-volume ratio of the local subvolume increases, which is
bad, since more communication needs to be done per unit of computation
(the communication effort typically increases with the surface area,
whereas the computing effort typically increases with the volume).  In
the fortunate case that effect (1) is dominant, one would observe
superlinear scaling.  More typically, effect (2) dominates, leading to
sublinear scaling or even saturation.

\begin{figure}
  \centering
  \includegraphics[height=30mm]{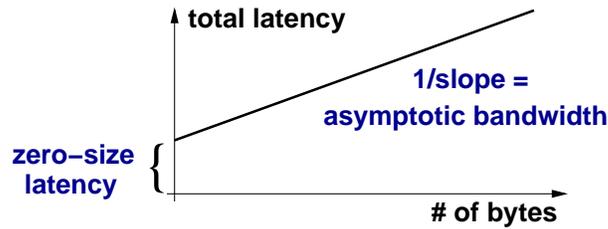}
  \caption{\label{fig:latency}In a simple model of network
    performance, the time (or total latency) it takes to transfer a
    packet of a certain size depends on the zero-size latency (which
    includes sender and receiver overhead as well as time of flight)
    and on the asymptotic (or maximum) bandwidth of the network: total
    latency = zero-size latency + (\# of bytes)/(asymptotic
    bandwidth).  Sometimes people quote the actual bandwidth (or
    throughput), defined as (\# of bytes)/(total latency).  For large
    packet size this converges to the asymptotic bandwidth.}
\end{figure}

A large part of the effort in designing a massively parallel machine
is directed at suppressing, or at least partially evading, the
consequences of effect (2).  The corresponding techniques are known
as ``communications latency hiding''.  To understand this better, we
remind the reader of two basic parameters characterizing a network
connection, i.e., bandwidth and latency, shown schematically in
Fig.~\ref{fig:latency}.
(In the remainder of this paper, we will use the terms latency and
bandwidth to denote the zero-size latency and the asymptotic
bandwidth, respectively, unless stated otherwise.)  If a large number
of bytes needs to be transfered, bandwidth is the more important
factor since the latency only makes a small contribution to the total
transmission time.  However, for small packets, latency is the
dominant part of the total time.  This is the relevant case when the
local volume becomes small, as it does in the case of strong scaling,
since the amount of information that needs to be communicated to
another node, and therefore the packet size, is small.  (A simple
example is provided by ping times in online gaming: the amount of
information to be transmitted, e.g., the target coordinates of the
enemy, is small, and thus the success of the gamer critically depends
on small latencies, i.e., short ping times.)  The term communications
latency hiding refers to techniques in which computation and
communication are overlapped, i.e., the CPU can do useful computations
while the communication is in progress.

Various aspects of network performance can be measured in special
benchmarks.  For example, the classic measure of latency and bandwidth
is a ``ping-pong'' benchmark in which a single message is bounced back
and forth between two nodes.  In a ``ping-ping'' benchmark, two
messages are bounced back and forth in both directions simultaneously,
and the measured performance gives an indication of the interference
between opposite directions (which can be due to hard- and/or software
reasons).

We close this section by introducing the concept of capability and
capacity machines.  Roughly speaking, capability is the ability of a
machine to finish a given (typically very difficult) calculation in a
certain amount of time, whereas capacity refers to the ability of a
machine to carry out a given workload (typically consisting of many
jobs) in a certain amount of time.  A given machine has both
attributes, but one often speaks of capability machines (if the
sustained performance remains high in the strong-scaling case) and of
capacity machines (which only scale weakly).  In lattice QCD, both
types of machine are used.  Capability machines are needed to generate
dynamical configurations with small quark masses in a single Markov
chain.  Capacity machines are suitable for most analysis tasks and for
scans of parameter space.

\section{Design considerations}
\label{sec:design}

There are essentially three possibilities to obtain a massively
parallel machine: one can purchase a commercial supercomputer, one can
build a machine out of commercial off-the-shelf (COTS) components, or
one can custom-design a machine.  In all of these cases, the
considerations that play a role in the design or purchase process are
similar.  They naturally fall into two categories, hardware and
software, which we shall discuss in this section.  Most importantly,
the design should be balanced in the sense that there are no
bottlenecks.  It does not make sense to spend money or design efforts
on a high-performance component which is then slowed down by other
components that cannot keep up with it.

Of course, the machine should be designed or selected such that
lattice QCD applications perform well on it.  The workhorse in
dynamical-fermions simulations is the conjugate gradient (CG)
inversion algorithm and variants thereof.  Its main two ingredients
are (a) (sparse matrix)$\times$vector multiplication and (b) global
sums.  Therefore the benchmarks presented in Sec.~\ref{sec:machines}
will concentrate on these two operations and on the overall
performance of the CG inverter for various Dirac operators.  (For an
earlier set of benchmarks presented at LATTICE 2003, see
Ref.~\cite{latfor}.)

On the hardware side, attention should be paid to the following
points.\\[-6mm] 
\begin{itemize}\itemsep-1mm
\item Processor.  This is the heart of the machine.  It should have
  high peak performance, a large amount of cache or on-chip memory,
  and fast interfaces to external memory and the network.  It should
  be programmable in one or more standard languages, and high-quality
  compilers should be available for it.  Lattice QCD code should
  obtain high sustained performance on a single node.
\item Memory system.  In addition to on-chip memory, each node should
  have a sufficient amount of external memory that can be accessed
  with low latency and high bandwidth.  On a small machine, the memory
  might be shared, i.e., all nodes can access a common memory.  On a
  very large machine, the memory will be distributed, i.e., the nodes
  need to send messages to exchange data.  Some machines provide a
  mixture of shared memory (on subpartitions of the machine) and
  distributed memory, see, e.g., Sec.~\ref{sec:altix}.  In all cases,
  cache coherence should be guaranteed in hardware.
\item Network.  As stated above, the network should enable packet
  transfers with low latency and high bandwidth.  As much of the
  latency as possible should be hidden, e.g., by direct memory access
  (DMA) hardware that can unburden the CPU.  Also, hardware
  acceleration for critical operations (such as global sums) could be
  included in a custom-designed machine.  The topology of the network
  also plays an important role.  Switched networks typically suffer
  from higher latencies, whereas mesh-based networks require more
  cables.  Finally, there should be sufficient bandwidth for I/O to
  disk.
\item Other issues.  Obviously, the price of the various components
  enters the above mentioned cost-effectiveness function.  Expenses
  for power and cooling need to be taken into account.  The packaging
  density (which, incidentally, may affect the network performance
  through cable lengths) and the resulting footprint of the machine
  may be an issue if space is restricted.  Last but not least, the
  machine should be reliable and easy to maintain.\\[-6mm]
\end{itemize}
On the software side, several factors come into play.\\[-6mm]
\begin{itemize}\itemsep-1mm
\item Operating system (OS).  It should provide all necessary services
  without hindering performance.  Often the best solution is a
  single-user, single-process OS.
\item Compilers.  If standard languages such as C/C++ or Fortran are
  supported, compilers should be available that produce correct and
  efficient code.  Ideally, they should be free as well (Gnu tools).
  In addition, automatic assembler generators such as BAGEL
  \cite{bagel} are desirable to increase the code performance without
  undue burden on the programmer.
\item System libraries.  Communication calls and other important
  functions such as parallel I/O are typically implemented in system
  libraries.  For example, on machines with distributed memory, the
  MPI (message-passing interface) library is often used to send
  messages, whereas on shared-memory machines, the OpenMP library can
  be employed to specify shared-memory parallelism.  It is important
  that these libraries be implemented efficiently, making use of the
  hardware features specific to the machine.
\item Application code. There are many code systems in use in the
  lattice QCD community, some of which are freely available
  \cite{milc,chroma,cps,fermiqcd}.  Such a code system should be easy
  to understand, easy to use, and easy to extend.  It should also be
  portable to as many popular architectures as possible.  Of course,
  it should also provide high performance, which can be achieved by
  the use of optimized kernels for critical code sections (often in
  assembler).\\[-6mm]
\end{itemize}
Regarding the last two points, the efforts of the USQCD software team
and their collaborators \cite{usqcd} to create and maintain a portable
and efficient software system for lattice QCD deserve special mention.

\section{Parallel architectures used for lattice QCD}
\label{sec:machines}

In this section, we describe a representative subset of the parallel
architectures used for lattice QCD simulations, including benchmarks
and cost-effectiveness.  Since our emphasis is on parallel computing,
we shall not discuss single-node performance in detail.  On all of the
machines discussed below, the critical kernels have been optimized in
assembler (including clever cache management) and obtain single-node
performance close to the theoretical maximum.

Also, we will not address the issue of reliability in detail because
reliable numbers beyond rough estimates or anecdotal evidence are hard
to come by.  All of the machines discussed below are reliable in the
sense that the mean time between failures is longer than typical
checkpointing intervals.

A note of warning: when computing the cost-effectiveness of a
particular architecture, a factor of two can pop up if, as in the case
of PC clusters, single-precision calculations are a factor of two
faster than in double precision.  Whether or not double precision is
really necessary depends on the application, and in many cases one can
certainly get by with a mixture of mainly single and some double
precision.  Having said this, the numbers quoted below are all based
on double precision unless stated otherwise.

\subsection{Commercial supercomputers}

Commercial machines are available from a variety of vendors such as
IBM, Cray, Hitachi, SGI, and others.  Typically, they are designed to
obtain reasonable sustained performance for a wide range of
applications.  This implies that their architecture is not necessarily
optimal for lattice QCD.  Furthermore, they are typically rather
expensive and can therefore be found mainly in large national or
regional computing centers where the lattice community can only get a
fraction of the computing time.  Nevertheless, they provide a
non-negligible fraction of the total computing power available to our
community and are particularly useful when employed as capacity
machines.

We have chosen to concentrate on two machines that will be used for
lattice QCD simulations in the next few years, the SGI Altix and IBM's
BlueGene/L.  For the performance of lattice QCD code on the Earth
Simulator, see Ref.~\cite{earthsim}.

\subsubsection{SGI Altix}
\label{sec:altix}

The SGI Altix is based on the Intel Itanium 2 processor.  This is a
64-bit processor based on the so-called EPIC (explicitly parallel
instruction computing) architecture, which in turn is an enhancement
of the VLIW (very long instruction word) concept.  The basic idea here
is for the compiler to group independent machine instructions in
so-called bundles which can then be executed in a single clock cycle.
This results in simpler hardware than in the case of a superscalar
RISC CPU, where the burden of recognizing dependencies and scheduling
instructions dynamically lies with the hardware.  Clearly, the quality
of the compiler, i.e., its ability to parallelize instructions
statically, is of utmost importance for the performance of the Itanium
architecture.

\begin{figure}
  \centering
  \includegraphics[height=60mm]{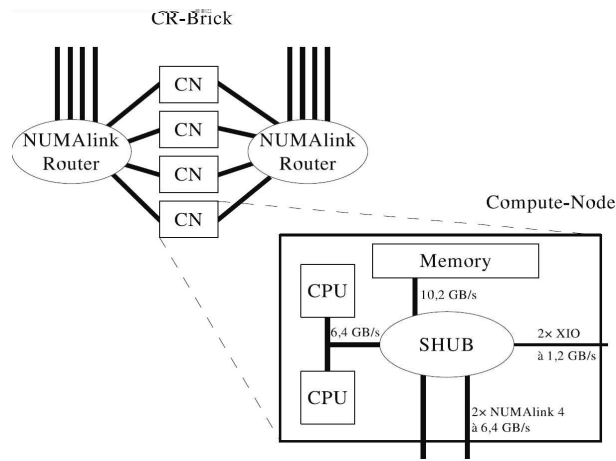}
  \caption{\label{fig:altix}Main components of the SGI Altix system.
    Figure taken from Ref.~\cite{altix}.}
\end{figure}

The architecture of the Altix is shown schematically in
Fig.~\ref{fig:altix}.
A compute node consists of two (single- or dual-core) CPUs that are
connected to local memory, to the network, and to I/O devices via a
custom-designed super-hub (SHUB) chip.  The maximum attainable
bandwidths are shown in Fig.~\ref{fig:altix}.  An obvious disadvantage
of this design is contention for memory and network accesses; however,
this is partially mitigated by the large cache size of the Itanium 2
(up to 9 MB).

The network of the Altix uses the ccNUMA (cache coherent non-uniform
memory access) architecture and is called NUMAlink.  Four compute
nodes are connected to form a compute-router (CR) brick by NUMAlink
routers in a fat-tree topology, as shown in Fig.~\ref{fig:altix}.  A
fat tree of CR bricks can then be built using first- and second-level
NUMAlink routers.  The MPI latency of NUMAlink4 is claimed to be 1
$\upmu$s, to which each hop across a router adds roughly 50 ns.  A
distinguishing feature of the Altix is the possibility of a
shared-memory domain with up to 512 CPUs.

The only lattice QCD benchmarks available so far are weak-scaling
studies of applications of the Wilson Dirac operator (Wilson Dslash in
short), i.e., the matrix$\times$vector component of the CG algorithm.
These benchmarks were performed by Thomas Streuer on the 64-node test
system in Munich, see Table~\ref{tab:altix}.  Note that the local
volume is so small that the problem fits in L3 cache.  All
communications were done using shared memory (shmem\_put), and
hand-coded assembler was used to obtain the quoted numbers.
\begin{table}
  \centering
  \begin{tabular}{c|c|c}
    \# CPUs & $V_\text{global}$ & sustained perf.\\[0.5mm]\hline
    &&\\[-4mm]
    8 & $8^3\times4$ & 31\%\\
    16 & $8^4$ & 26\%\\
    32 & $8^3\times16$ & 30\%\\
    64 & $8^3\times32$ & 28\%
  \end{tabular}
  \caption{\label{tab:altix}Performance of the Wilson Dslash on the
    SGI Altix. In all cases, $V_\text{local}=4^4$.}  
\end{table}
The performance is quite impressive, but it remains to be seen how it
scales as the global sums are included and as the number of nodes is
increased, once the full machine is available.

At the Leibniz Computing Center in Munich, a system with 33 TFlop/s
peak (2,560 1.6 GHz dual-core Montecito sockets with 6 MB L3 cache
each and 20 TB of main memory) will be operational at the beginning of
2006, and an upgrade to 69 TFlop/s peak (3,328 2.6 GHz dual-core
Montvale sockets with 9 MB L3 cache each and 40 TB of main memory)
will occur in 2007.  The purchase price for this system is 38 million
Euro, not including substantial running costs.  Assuming that the
performance numbers quoted in Table~\ref{tab:altix} go down on larger
machines by a factor of two or so, and figuring in the additional cost
of global sums, an optimistic estimate for the cost-effectiveness of
this particular machine for lattice QCD is thus on the order of
5 Euro per sustained MFlop/s in 2007, i.e., two years from now.  As
we shall see, this is substantially more expensive than the 2005
numbers for the other machines to be discussed below.

\subsubsection{BlueGene/L}
\label{sec:bgl}

IBM's flagship supercomputer, BlueGene/L (BG/L in short), currently
occupies the No.~1 and 2 spots on the Top 500 list (183 TFlop/s peak
at LLNL and 115 TFlop/s peak at IBM Watson).  Its design has been
influenced by, and is similar to, that of QCDOC (and QCDOC's
predecessor, QCDSP), see Sec.~\ref{sec:qcdoc} below.

The architecture of BG/L has been described in detail in
Ref.~\cite{bgl1}, the performance of lattice QCD code was addressed in
Ref.~\cite{bgl2}, and an entire journal issue was dedicated to BG/L in
Ref.~\cite{ibm}.  We briefly summarize the relevant points here.  The
heart of the machine is an application-specific integrated circuit
(ASIC) containing two 700 MHz PowerPC 440 CPU cores to each of which
are attached two 64-bit FPUs that can be utilized to execute two FMA
instructions per clock cycle.  The peak performance of a single ASIC
is thus 5.6 GFlop/s.  There is a shared 4 MB L3 cache on chip, and
external memory is distributed, with 512 MB per node.  The nodes are
connected in a three-dimensional torus with nearest-neighbor
connections (but without DMA capabilities).  In addition, there is a
global tree network that is used for global reductions.

\begin{figure}
  \centering
  \includegraphics[height=60mm]{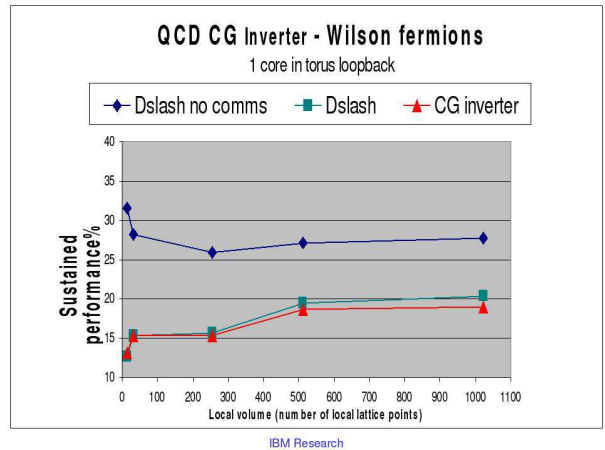}
  \caption{\label{fig:bgl}Strong scaling of Wilson Dslash and CG on
    BlueGene/L~\cite{bgl2}.  Source: Pavlos Vranas.}
\end{figure}

BG/L can be run in one of two modes.\\[-6mm]
\begin{itemize}\itemsep-1mm
\item Co-processor mode.  One of the 440 cores is used for
  computation, the other one for communication.  Thus, the peak
  performance is cut in half to 2.8 GFlop/s.
\item Virtual-node mode.  Both of the 440 cores are used for
  computation and communication.  The peak performance is now the full
  5.6 GFlop/s, but computation and communication cannot be
  overlapped.\\[-6mm] 
\end{itemize}
Which of these two modes results in higher sustained performance
depends on the application.  The performance numbers quoted in
Ref.~\cite{bgl2}, and reproduced in Fig.~\ref{fig:bgl} for
convenience, were obtained in virtual-node mode using C/C++ inline
assembler and carefully tuned code that takes advantage of special
hardware features of BlueGene/L.
Although the numbers in Fig.~\ref{fig:bgl} were obtained on only one
node (i.e., two cores), the performance estimates are realistic since
the torus network was used for the communications.  A weak-scaling
benchmark of the Wilson CG (using a local volume of $4^3\times16$)
showed a very mild degradation of the sustained performance as the
number of nodes was increased~\cite{bgl2}.  This was most likely due
to the fact that the global sums were done on the torus instead of the
global tree network because the necessary software was not yet
available at the time.  They can now be done on the global tree, so
the degradation should disappear.

For reasons unbeknownst to the author, the groups that bought a BG/L
machine are not allowed to disclose the purchase price.  Of course,
rumors abound, and for the sake of argument we will estimate the price
to be \$2 million per rack.  A rack consists of 1,024 nodes, thus the
sustained performance of a large-scale lattice QCD application
(assuming, say, 17\% of peak) is on the order of 1 TFlop/s per rack,
which translates into a cost-effectiveness of about \$2 per MFlop/s.

\subsection{PC clusters}

PC clusters have been reviewed at previous LATTICE conferences
\cite{lippert,holmgren04} and were presented this year as well
\cite{holmgren05}.  Clearly, they are a sensible choice for many
(lattice and non-lattice) groups.  The high-volume market for PCs
drives down the cost of the components so that small- or medium-size
clusters can be obtained by groups that do not have access to
multi-million dollar funding.  Also, the software running on PC
clusters is reasonably standard and can be used with relatively modest
effort.  

In the sub-TFlop/s range, PC clusters are probably the best overall
choice in terms of cost-effectiveness and ease of use.  The main
question is whether the available networking components allow PC
clusters to scale beyond this performance at a price competitive with
the custom-designed machines discussed in Sec.~\ref{sec:custom} below.
This is the point on which we will try to shed some light.  Note,
however, that PC clusters are very much a moving target since new
hardware appears on the market all the time.  Consequently, we can
only provide a snapshot of the current status.  For more details, we
refer to Refs.~\cite{edwards,holmgren}.  The currently installed
lattice QCD clusters with a peak performance of more than 1 TFlop/s
are summarized in Table~\ref{tab:clusters}.
\begin{table}
  \centering
  \begin{tabular}{c|c|c|c|c}
    Name & Institution & CPU & Network & Peak (TFlop/s)
    \\[0.5mm]\hline 
    &&&& \\[-4mm]
    ALICEnext & Wuppertal & 1024 Opteron & Gig-E (2d mesh) & 3.7\\
    4G & JLAB & 384 Xeon & Gig-E (5d mesh) & 2.2\\
    3G & JLAB & 256 Xeon & Gig-E (3d mesh) & 1.4\\
    Pion & Fermilab & 260 (520) P4-640 & Infiniband & 1.7 (3.4)\\
    W & Fermilab & 256 P4E-Xeon & Myrinet & 1.2
  \end{tabular}
  \caption{\label{tab:clusters}Existing lattice QCD clusters with more
    than 1 TFlop/s peak performance (double precision).  The Pion
    numbers in parentheses refer to a planned expansion in the fall of
    2005.} 
\end{table}

The hardware components making up a PC cluster can be roughly divided
into CPU, memory system, and network (including I/O interface), which
we shall discuss in turn.  Of the currently popular processors,
Intel's Pentium 4 and Xeon as well as AMD's Opteron appear to offer
the best price-performance ratio.  The difference between the P4 and
the Xeon is that the former is strictly a uniprocessor whereas the
latter is capable of symmetric multi-processing (SMP).  The Opteron
has the advantage of having the memory controller on chip, but it
tends to be a little pricier.  Since most of the available benchmarks
were obtained with Intel CPUs, we will concentrate on these chips.
While there are some differences between the various processors on the
market, the issues that arise in terms of cluster building are very
similar for all of them, so the conclusions we will draw are
universal.

\begin{figure}
  \centering
  \includegraphics[height=60mm]{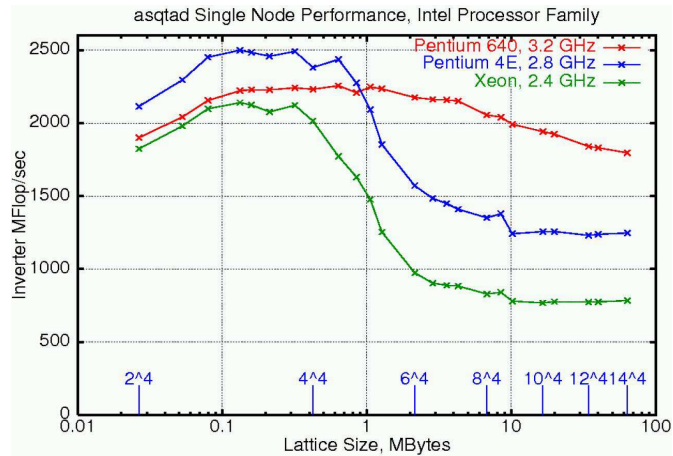}
  \caption{\label{fig:pc_single}Single-node performance of the asqtad
    inverter (in single precision) on various Intel CPUs.  Cache size
    and frequency of the front-side bus are as follows.  Xeon: 0.5 MB,
    400 MHz; P4E: 1.0 MB, 800 MHz, P640: 2.0 MB, 800 MHz.  From
    Ref.~\cite{holmgren}.}
\end{figure}

Clock frequencies nowadays are approaching 4 GHz so that the peak
performance of a single processor is getting close to 8 GFlop/s.
Using the SIMD graphics vector extensions provided by Intel and AMD
processors \cite{csikor,luescher}, lattice QCD code can obtain 50\% or
more of peak when operating from cache \cite{pochinsky,holmgren04},
see Fig.~\ref{fig:pc_single} for a typical benchmark.
Once the problem moves out of cache, the performance is no longer
limited by the floating-point capabilities of the processor but by the
performance of the memory bus.  This can also be seen in
Fig.~\ref{fig:pc_single}.  Thus, if a physics problem is to be solved
on a small cluster, where the local volume tends to be large, the
memory bus is likely to be the bottleneck.  On a larger cluster, where
the local volume might fit in cache, the bottleneck moves to the
network interface.  

While newer processors typically come with a faster memory bus for
little or no extra money, the cost of better networking components can
represent a significant fraction (perhaps one half) of the total
expense for a PC cluster.  The relevant parameters are latency and
bandwidth, and the topology of the network plays a role as well.
Mesh-based networks with nearest-neighbor connections are well-suited
for lattice QCD and are quite cheap since switches are not necessary;
however, they require more cables, are less fault-tolerant, and
accumulate latencies when computing global sums.  Switched networks
provide all-to-all communication; however, nearest-neighbor
communications incur the switch latency, and low-latency switches are
expensive.

Popular network choices include Gigabit Ethernet (Gig-E), Myrinet,
Quadrics, and Infiniband.  The latter three are switched, whereas
Ethernet can be both switched or meshed.  All four choices provide
sufficient bandwidth for typical lattice QCD applications (if
necessary, multiple network cards can be used).  Latency is the more
difficult problem, and the rule of thumb is that lower latency is
reflected in a much higher price.  Since strong scaling needs low
latencies as explained above, this is the main issue determining
whether PC cluster are competitive (in terms of cost-effectiveness)
for large-scale applications.

Before presenting numbers, we need to mention two more points that are
relevant to this issue.  First, the messages sent through the network
typically go via the PCI bus, and one has to make sure that this is
not the bottleneck.  For example, PCI-X is sufficient for Gig-E but
throttles Infiniband performance, whereas PCI-Express is unlikely to
limit the network performance for the next few years.  (In addition,
sub-optimal PCI chipset implementations may set further limits on the
performance.)  Second, an efficient software implementation of the
communication calls is essential to obtain low latencies.  For
example, TCP/IP has a high software overhead, whereas high-performance
libraries (such as the M-VIA cluster communications library
\cite{mvia} and the QMP message-passing library \cite{qmp}) are much
leaner and therefore result in much lower latencies.  Using these lean
libraries and PCI-Express, typical latencies are on the order of
12~$\upmu$s for Gig-E, 5~$\upmu$s for Myrinet, and 3.5~$\upmu$s for
Infiniband.  An interesting future alternative (for systems with a
Hypertransport interface) is Infinipath with latencies below
1~$\upmu$s.

\begin{figure}
  \centering
  \includegraphics[height=60mm]{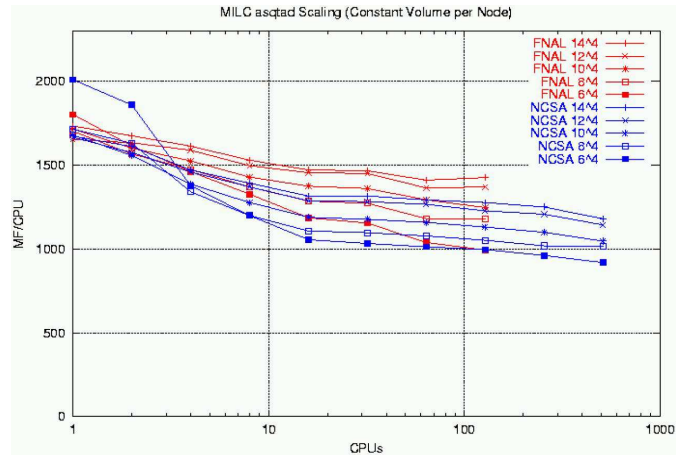}
  \caption{\label{fig:pc_asqtad}Performance of the MILC asqtad
    inverter on the T2 cluster (MILC code v6) and on the Pion cluster
    (MILC code using QDP, optimized by James Osborn).  From
    Ref.~\cite{holmgren}.} 
\end{figure}

\begin{figure}
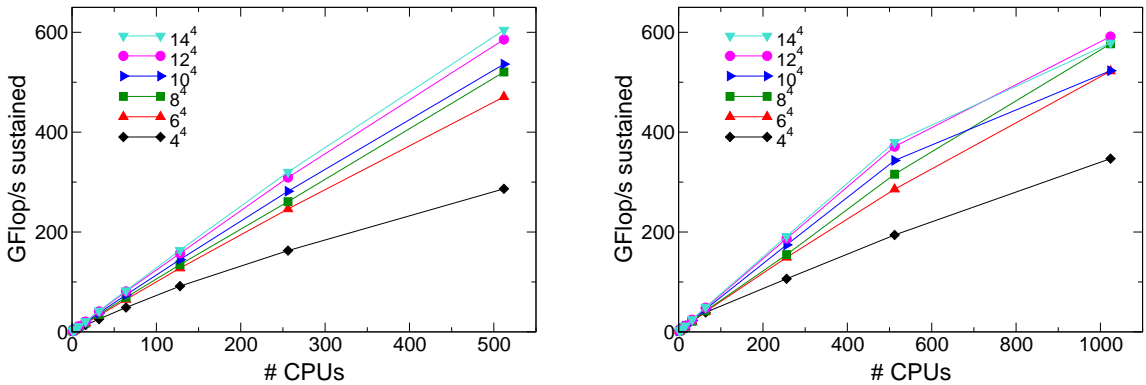

  \centering
  \vspace*{1mm}
  \includegraphics[height=50mm]{plots/t2_vmi_single_total.eps}\hfill
  \includegraphics[height=50mm]{plots/t2_vmi_dual_total.eps}
  \caption{\label{fig:pc_dual}Performance of the MILC asqtad inverter
    on the T2 cluster using only a single CPU per motherboard (left)
    and using both CPUs (right) \cite{steve}.}
\end{figure}

\begin{figure}
  \centering
  \vspace*{5mm}
  \includegraphics[height=50mm]{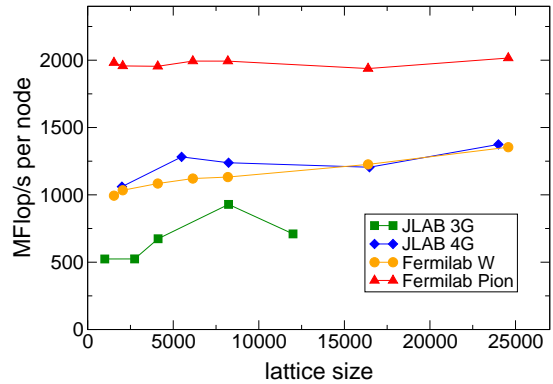}
  \caption{\label{fig:pc_dwf}Weak scaling of the DWF inverter on four
    different PC clusters \cite{don}.  The peak performances per node
    in MFlop/s are 5300 (3G), 5600 (4G), 4800 (W), and 6400 (Pion).
    The labels on the horizontal axis correspond to the global
    4-dimensional lattice size $V_4$, and the 5-dimensional lattice
    size is $V_4\times16$.}
\end{figure}

Of course, the real question is how these numbers translate into
lattice QCD application code performance and scalability.  To answer
this question, we present benchmarks obtained on the PC clusters at
Fermilab and Jefferson Lab.  (Note that in all cases in this section,
single-precision performance is plotted, whereas the peak performance
is quoted for double precision.)  Figure~\ref{fig:pc_asqtad} shows the
performance of the MILC asqtad inverter on the Pion cluster at
Fermilab and on the T2 cluster at the NCSA (the latter consists of 512
dual 3.6 GHz Xeon processors connected via Infiniband over PCI-X).
The curves are weak-scaling curves since the local volume is constant
as the number of nodes is increased, but since several local volumes
were used the plot also gives an indication of the strong-scaling
behavior.  We see that going from $14^4$ to $6^4$ on machines with up
to 512 nodes decreases the sustained performance by about 30\%.  We
also see that the Pion cluster outperforms the T2 cluster even though
its processors are slower.  This is due to the use of PCI-Express
instead of PCI-X and to the more efficient software used on Pion.
Two more lessons can be drawn from asqtad inverter benchmarks on the
T2 cluster displayed in Fig.~\ref{fig:pc_dual} \cite{steve}.
First, once the local volume is as small as $4^4$, the sustained
performance drops significantly.  Thus, even with an expensive,
low-latency network such as Infiniband, strong scaling is limited.
(To be fair, a local volume of $6^4$ is quite reasonable for most
problems.)  Second, the performance is roughly the same whether a
single or both CPUs are used, i.e., the second Xeon is essentially
useless in this case.  This is due to the fact that with the chipset
used on that cluster, the memory bandwidth is already saturated by a
single CPU.  (The Opteron with its memory controller on the chip would
not have this problem.)  Finally, weak-scaling curves for the
domain-wall fermion (DWF) inverter (written in assembler by Andrew
Pochinsky) are presented in Fig.~\ref{fig:pc_dwf} \cite{don}.
The observation here is that with roughly comparable single-node
performance, the Infiniband cluster obtains much higher sustained
performance.  

What processor/network combination provides the best price-performance
ratio is a function of time.  The US cluster community currently
considers P4/Infiniband to be optimal.  Other groups might
(justifiably) have different preferences, see, e.g., the Opteron/Gig-E
solution in Wuppertal.  The 4G cluster at JLAB currently sustains 650
GFlop/s on the DWF inverter with a local lattice size of
$8^4\times16$, which translates into a cost-effectiveness of \$1.1
per sustained MFlop/s in single precision (twice that for double
precision).  In the future, one hopes to sustain 1-2 TFlop/s on
$\sim$1,000 nodes at a cost-effectiveness of \$1 per sustained
MFlop/s (single precision).  The additional cost for power and cooling
is estimated to be about 5\% of the purchase price per year.

\subsection{Custom-designed machines}
\label{sec:custom}

As mentioned in Sec.~\ref{sec:basics}, lattice QCD is well suited for
parallelization.  It is possible to design special-purpose capability
machines that make use of the simplifying features of lattice QCD to
obtain superior scalability, i.e., high sustained performance on a
very large number of nodes, at very low cost.  (The Grape
project~\cite{grape} is an extreme example of this idea in another
field, i.e., astrophysics.)  This approach is sensible only if the
overall hardware budget available to the lattice community is large
enough so that the higher cost-effectiveness of such machines is not
spoiled by the development costs.  Fortunately, this is true, and
therefore special-purpose machines have been and will continue to be
developed.  Most of the history of these machines is described in the
Machines \& Algorithm section of the yearly LATTICE proceedings.
Here, we will concentrate on the two latest machines, apeNEXT and
QCDOC.

\subsubsection{apeNEXT}

\begin{figure}
  \centering
  \includegraphics[height=50mm]{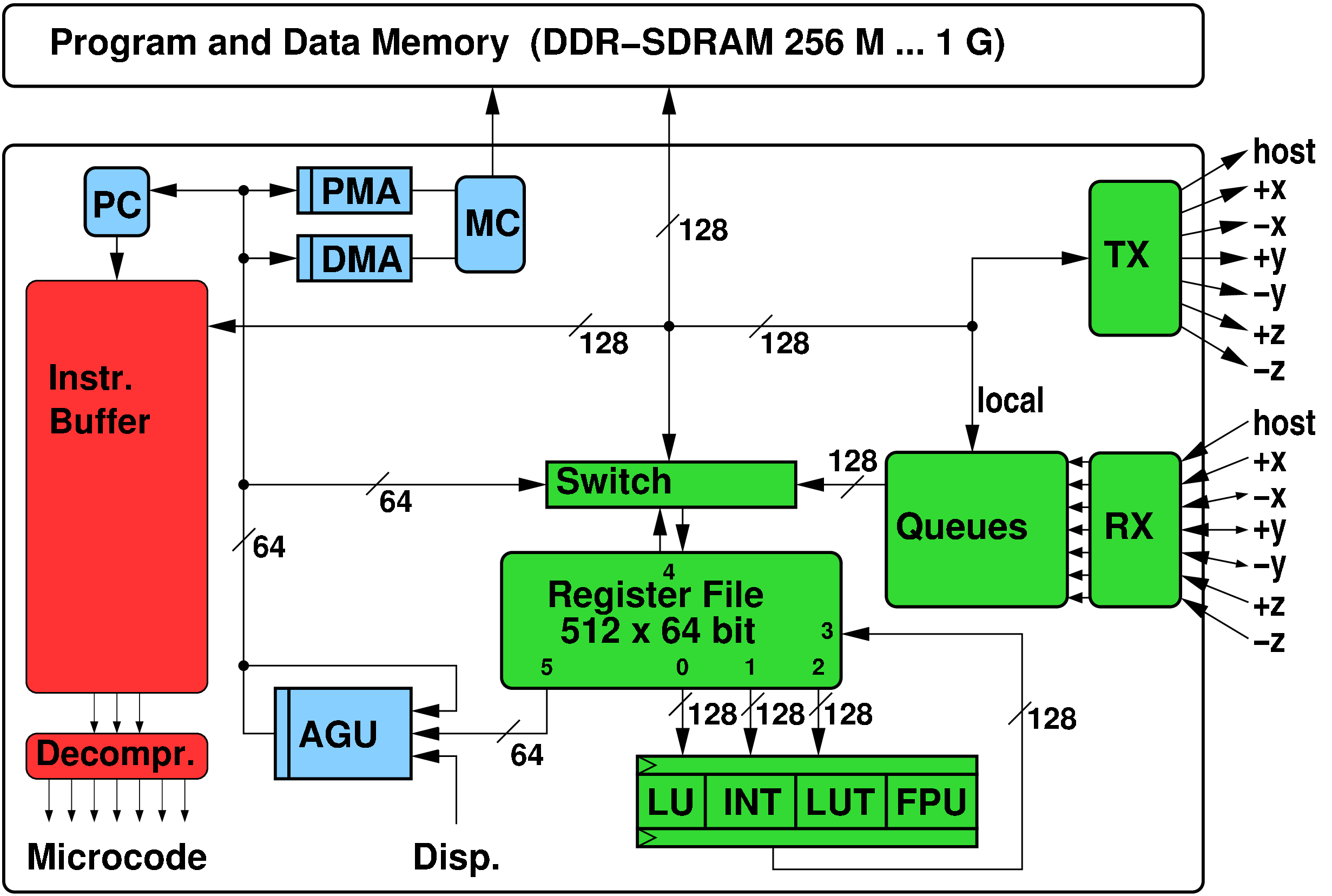}
  \caption{\label{fig:jt}Block diagram of the J\&T processor, the
    basic building block of apeNEXT.  Taken from Ref.~\cite{apenext}.}
\end{figure}

The apeNEXT computer was designed and developed by a collaboration of
INFN Ferrara and Rome, DESY Zeuthen and the Universit\'e de Paris-Sud,
Orsay.  It is the successor of earlier APE machines (APE, APE100,
APEmille) and was presented at last year's
conference~\cite{apenext:lat04}.  Further details can be found in
Ref.~\cite{apenext}.  The heart of the machine is a custom ASIC, the
J\&T processor, whose FPU can execute 8 Flops (a complex $a\times b+c$
operation) per clock cycle.  Running at 160 MHz, this corresponds to a
peak performance of 1.3 GFlop/s.  The apeNEXT chip is shown in
Fig.~\ref{fig:jt}.
In addition to the FPU, it contains a 64 kB instruction buffer (as in
the case of the Itanium, see Sec.~\ref{sec:altix}, the VLIW idea is
used), 10 kB of prefetch queues, a 4 kB register file, a controller for
external memory with a bandwidth of 2.6 GB/s, and the communications
hardware with DMA capability.  The topology of the network is a
three-dimensional torus with nearest-neighbor connections, allowing
concurrent send and receive along any of the directions with a
bandwidth of about 125 MB/s per link and direction.

As for the physical design of the machine, 16 daughterboards
containing one ASIC each are mounted on a so-called processing board,
which in addition contains an FPGA (field-programmable gate array) for
the handling of global signals and an I2C interface.  A backplane
holds 16 processing boards, and a rack consists of two stacked
backplanes, i.e., of 512 nodes.  The footprint of a rack is less than
1 m$^2$, and the power consumption has been measured to be about 8.5
kW per rack.

apeNEXT can be programmed in TAO (a Fortran-like language specific to
APE machines), in C (with an lcc-based compiler written by the apeNEXT
team), or in SASM (high-level assembler).  The TAO and C compilers are
stable, but work is ongoing to improve code efficiency.

\begin{figure}
  \centering
  \vspace*{1mm}
  \includegraphics[height=52mm]{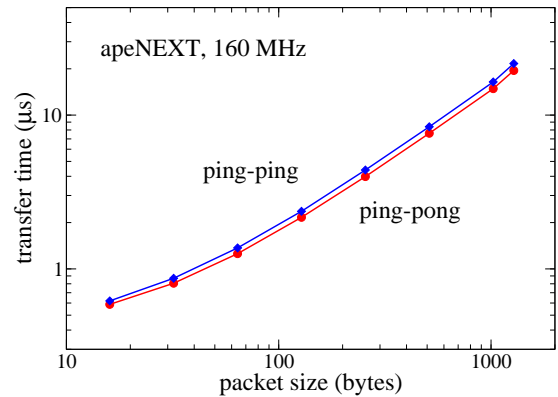}
  \caption{\label{fig:ape-ping}Measurements of the total ping-ping and
    ping-pong latencies as a function of the packet size for a single
    bidirectional link of apeNEXT at 160 MHz \cite{hubert}.  The
    benchmarks were written in SASM, use only registers and run in
    cache.}
\end{figure}

Unfortunately, due to an error made by the chip manufacturer, the
apeNEXT project was somewhat delayed so that application code
benchmarks could not be obtained yet on large machines.  Currently,
single-node performance is 54\% of peak for the Wilson Dslash
(hand-coded in assembler) and 37\% for the TAO-based clover CG (the
latter number is expected to go up with further optimization).  A
benchmark of the apeNEXT network performance is displayed in
Fig.~\ref{fig:ape-ping}, where the total ping-ping and ping-pong
latencies for a single bidirectional link are plotted as a function of
the packet size.
As can be seen from the figure, the ping-pong latency of the network
is at least 500 ns, but most of this can be hidden, except in the case
of global sums.  A global sum takes $2(N_x+N_y+N_z-3)$ steps of $\sim$
30 cycles each on an $N_x\times N_y\times N_z$ processor mesh, so on
1,024 nodes a global sum takes about 11 $\upmu$s.  Strong-scaling
numbers do not exist yet for the reasons mentioned above, but the
machine should perform close to linear scaling since (a) the
communications overhead for the Wilson Dslash is only 4\% on a local
volume of $2^3\times16$ and (b) the global sums have no significant
impact on the performance on machines with up to 4,096 nodes.

A short update on the progress of apeNEXT: a 512-node and a 256-node
prototype rack using version A of the ASIC are running stably, and two
512-node racks using version B of the ASIC have been assembled and
tested, with a target frequency of 160 MHz.  The physics production
codes are running with almost no modifications with respect to
APEmille, but further optimization is needed to reach the efficiency
of the benchmark kernels.

The following installations of apeNEXT are planned: 12 racks INFN, 6
racks Bielefeld, 3 racks DESY, and 1 rack Orsay.  Based on a 160 MHz
clock rate, the peak performance of a rack is 0.66 TFlop/s.  The price
is 0.60 Euro per peak MFlop/s \cite{lele}, so the cost-effectiveness
is about 1.2~Euro per sustained MFlop/s.

\subsubsection{QCDOC}
\label{sec:qcdoc}

\begin{figure}
  \centering
  \includegraphics[height=60mm]{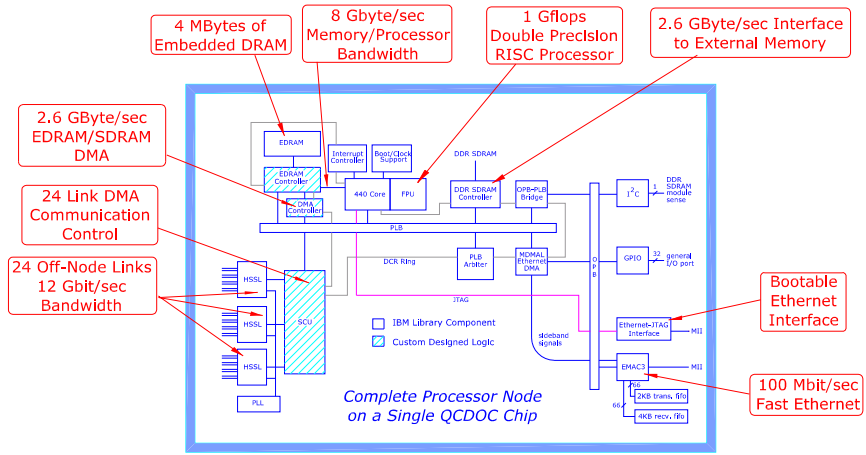}
  \caption{\label{fig:asic}Block diagram of the QCDOC ASIC.  Note that
    the peak performance and the memory and network bandwidths are
    based on a 500 MHz clock.}
\end{figure}

The QCDOC (QCD On a Chip) computer is a successor of the QCDSP machine
\cite{qcdsp} and was developed by a collaboration of Columbia
University, the UKQCD collaboration, the RIKEN-BNL Research Center,
and IBM Research.  It has been discussed in detail at last year's
conference \cite{qcdoc:lat04}, and further material can be found in
Ref.~\cite{qcdoc}.  The architecture of the machine is based on
the QCDOC ASIC, shown schematically in Fig.~\ref{fig:asic}.
It contains a number of elements from IBM's technology library (e.g.,
PowerPC 440 embedded CPU core, 64-bit FPU with one FMA per cycle, 4~MB
embedded DRAM, memory controller, Ethernet controller, and high-speed
serial links for the communications network) as well as
custom-designed logic optimized for lattice QCD (e.g., serial
communications unit with DMA capability, prefetching EDRAM controller,
and bootable Ethernet-JTAG interface).  Running at a conservative
clock frequency of 400 MHz, its peak performance is 0.8 GFlop/s.  The
network has the topology of a six-dimensional torus with
nearest-neighbor connections.  As in the case of apeNEXT, concurrent
send and receive is possible along any of the directions.  The extra
dimensions can be used to divide the machine into smaller partitions
in software.

Two ASICs are mounted on a daughterboard, and 32 daughterboards are
mounted on a motherboard.  The motherboards can either be put in an
air-cooled crate (8 each) or in a water-cooled cabinet (16 each).  The
footprint of a crate or a cabinet is about 1 m$^2$, and the power
consumption is about 8 kW per cabinet.

There are currently five QCDOC installations: 14,720 nodes at
Edinburgh (UKQCD), 14,140 nodes at BNL (USQCD), 13,308 nodes at the
RIKEN-BNL Research Center, 2,432 nodes at Columbia University, and 448
nodes at the University of Regensburg.  

QCDOC can be programmed in C, C++, or PowerPC assembler.  There are
two sets of compilers available, the free gcc/g++ and IBM's commercial
xlc/xlC.  The latter produce faster code, but since critical kernels
are available in assembler the GNU tools are generally sufficient.

\begin{figure}
  \centering
  \includegraphics[width=60mm,angle=270]{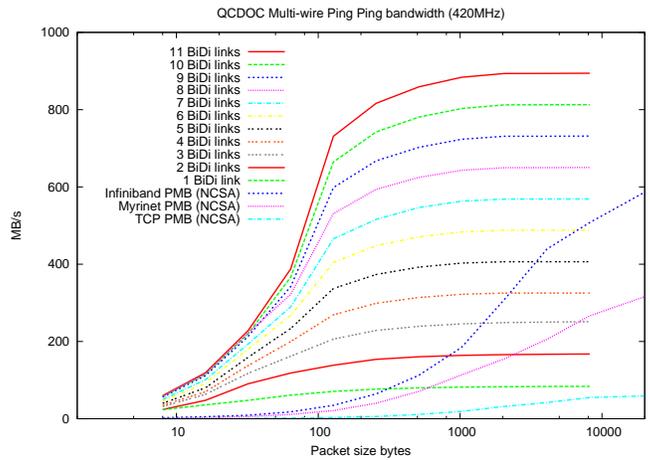}
  \caption{\label{fig:qcdoc_bw}Measurements of the actual
    bidirectional bandwidth obtained on the QCDOC network at 420 MHz
    as a function of the message size and the number of active links.
    Also shown are the bandwidths obtained with three commercial
    networks (Infiniband, Myrinet, and Gbit Ethernet).}
\end{figure}

We first present two benchmarks measuring the network performance of
QCDOC.  (Most of the benchmarks presented in this section were
obtained by P.A. Boyle, with the exception of the asqtad performance
in Table~\ref{tab:qcdoc}, which was obtained by C. Jung.)
Figure~\ref{fig:qcdoc_bw} shows the aggregate actual bidirectional
bandwidth obtained as a function of the message size and the number of
active links.
The maximum bidirectional bandwidth (including protocol overhead) at
420 MHz is 84 MB/s per link.  Two points to note are that the
multi-link bandwidth is comparable to the memory bandwidth and that a
single link obtains 50\% of the maximum bandwidth already on very
small packets (32 bytes).
\begin{figure}
  \centering
  \includegraphics[width=60mm,angle=270]{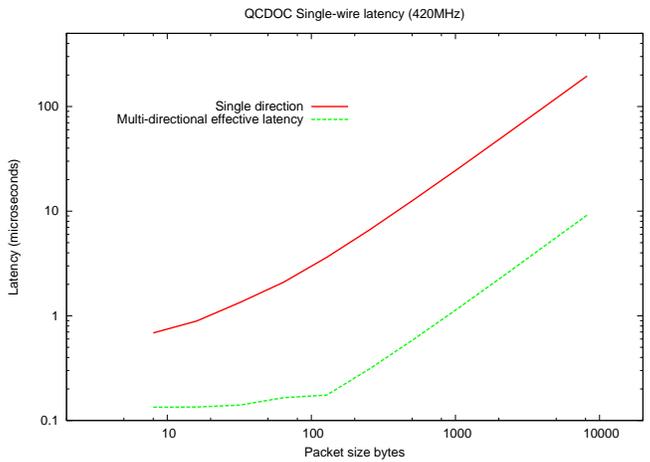}
  \caption{\label{fig:qcdoc_lat}Total ping-ping latency for a single
    direction and multi-directional effective latency obtained on the
    QCDOC network at 420 MHz as a function of the message size.}
\end{figure}
Figure~\ref{fig:qcdoc_lat} shows the total ping-ping latency at 420 MHz
for a single link as a function of the message size.  Also shown is
the multi-directional effective latency, which is obtained by
transmitting in 12 directions simultaneously and dividing the transfer
time by 12.

\begin{figure}
  \centering
  \includegraphics[width=60mm,angle=270]{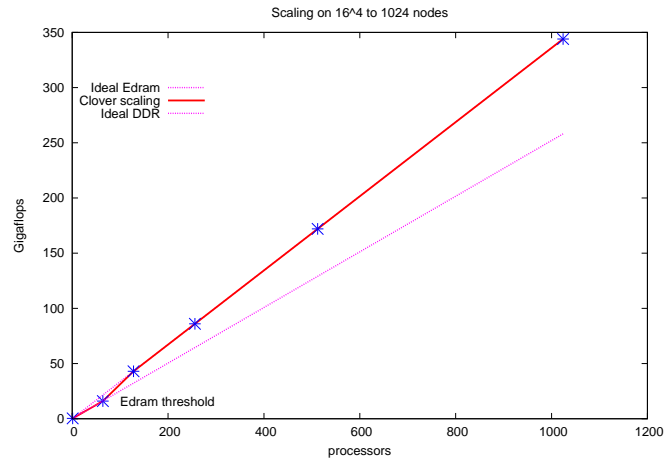}
  \caption{\label{fig:qcdoc_scaling}Strong-scaling benchmark for the
    sustained performance of the clover conjugate gradient on QCDOC at
    420 MHz.  The global volume is $16^4$.}
\end{figure}

The network performance of QCDOC is reflected in the strong-scaling
benchmark shown in Fig.~\ref{fig:qcdoc_scaling}.
Also shown are the ideal (i.e., linear) scaling curves when the data
are located in on-chip (EDRAM) or off-chip (DDR) memory.  We observe
superlinear scaling when the problem moves from DDR to EDRAM, and
linear scaling afterwards.  On 1,024 nodes, the global volume of
$16^4$ used in this benchmark corresponds to a local volume of
$2^2\times4^2$.  The same sustained performance can be expected on
16,384 nodes using a $32^4$ global volume.

The sustained performance of various lattice QCD applications at 420
MHz for a local volume of $4^4$ is displayed in Table~\ref{tab:qcdoc}.
\begin{table}
  \centering
    \begin{tabular}{c|c|c|c}
      Action & \# Nodes & matrix$\times$vector & CG
      performance \\[0.5mm] \hline 
      &&& \\[-4mm]
      Wilson & 512 & 44\% & 39\% \\
      Asqtad & 128 & 42\% & 40\% \\
      DWF & 512 & 46\% & 42\% \\
      Clover & 512 & 54\% & 47\%
    \end{tabular}
    \caption{\label{tab:qcdoc}Lattice QCD application code performance
      on QCDOC in percentage of peak for a local volume of $4^4$.}
\end{table}
Additional results can be found in the second paper of
Ref.~\cite{qcdoc:lat04}.  The author was unable to obtain updated
benchmarks on larger machines since those machines are busy producing
physics results.  There are currently three RHMC jobs (with different
quark masses) running on 4,096 nodes each in the US and the UK.  The
sustained performance of these jobs is only about 35\% of peak since
(a) the local volume of $6^3\times2\times8$ does not fit in EDRAM and
(b) the linear algebra required for the multi-shift solver is running
from external memory.  Nevertheless, the total sustained performance
of each of these jobs is over 1.1 TFlop/s.  The two-flavor DWF
inverter, which is part of the RHMC algorithm, sustains 40\% of peak.
When going from 1,024 to 4,096 nodes, superlinear scaling was observed
since part of the problem moved into EDRAM and since there was no
noticeable degradation from the communications overhead.

A 12-rack machine consisting of 12,888 nodes has a peak performance of
10 TFlop/s at 400 MHz.  The price is \$0.45 per peak MFlop/s
\cite{nhc}, resulting in a cost-effectiveness of about \$1.1 per
sustained MFlop/s.  The additional cost for power and cooling is less
than 2\% of the purchase price per year.

\section{Speculations on future machines}
\label{sec:future}

Parallelization is the wave of the future, and not just in the field
of high-performance computing.  The increased power density and the
associated heat dissipation problem is a major issue in the chip
manufacturing industry, which sets a limit on the clock frequencies
that can reasonably be achieved under normal operating conditions.
The way out of this problem, which is already being taken by most of
the major players, is to put multiple CPU cores on a single chip.
While one could in principle run independent jobs on these cores, this
will become increasingly difficult as the number of cores per chip
increases since (a) the memory (and network) bandwidth will not be
sufficient for many independent jobs and (b) the total amount of
memory per chip would have to be very large (e.g., if we assume that a
typical application requires 1 GB of memory, a chip with 32 cores
would require 32 GB of memory, which is not just a technological
challenge but also prohibitively expensive).  On-chip parallelization
would overcome these problems since the memory requirements per core,
in terms of both bandwidth and amount, would be correspondingly
smaller.  Hardware designers dream of automatic parallelization (i.e.,
the detection of dependencies) in hardware, similar to the idea of
dynamic instruction scheduling in superscalar RISC CPUs, see
Sec.~\ref{sec:altix}.  However, it is not clear how practicable this
approach is, and therefore a more conservative approach is to adjust
the programming models to make efficient use of multi-core chips.
What is necessary is more fine-grained parallelism on the chip.  To
get an idea of what is meant by this phrase, imagine a situation where
the local volume per core is less than one site, i.e., the loops
associated with a single site are distributed among several cores.
Application code dealing with such a situation would probably contain
a mixture of pthreads, OpenMP, and MPI (or similar implementations of
the corresponding concepts), and data dependencies would have to
specified by hand (e.g., in the form of compiler directives as in the
case of OpenMP).  The challenge to the system programmer is to write
libraries that hide these details from the general user, and it will
be interesting to see how this challenge is met in the future.  Of
course, lean software is essential to obtain as much performance as
possible from such multi-core chips.

Will custom-designed chips continue to play a role lattice QCD?  One
of the problems facing chip designers are the very high NRE
(non-recurring engineering) costs of large ASICs, which can be on the
order of a few \$$10^6$ or so.  This is not a life-or-death issue for
chips that will be manufactured in large volumes.  However, for a
dedicated lattice QCD chip of which only a few 10,000 units will be
made, the NRE costs would represent a non-negligible fraction of the
total cost.  In addition, the financial risk associated with a
possible re-spin of the chip would be very high.  Some of the
possibilities to deal with this issue are (a) to collaborate with a
big company (such as IBM in the case of QCDOC), (b) to use FPGAs in
the development phase, or (c) to combine a commercial processor with a
custom-designed network ASIC (as in the case of QCDSP).  (For an
implementation of the Wilson-Dirac operator on an FPGA, see
Ref.~\cite{callanan}.)

There is no doubt that PC clusters will contribute a major percentage
of the computing power available for lattice QCD.  Since PC clusters
are used by many communities, the market is sufficiently large to
drive improvements in cluster hardware at all fronts (processors,
memory bus, chipsets, network interfaces, etc.).  For example, a
custom-designed three-dimensional interconnect architecture for PC
clusters called APENet was presented at this conference \cite{apenet}.
A continuous software effort is necessary to get maximum performance
out of the new architectures.  It will be interesting to see how
multi-core chips perform in a cluster environment.

Let us mention concrete proposals for new machines and speculate on
some possible future projects.\\[-6mm]
\begin{itemize}\itemsep-1mm
\item PC cluster upgrades are planned at JLAB (a clone of the 260-node
  Fermilab Pion cluster will go online in the spring of 2006) and at
  Fermilab (an Infiniband cluster with $\sim$1,000 processors, either
  $\sim$500 dual Xeons or $\sim$1,000 single Pentiums depending on
  price/performance at the time, will be installed in the fall of
  2006) \cite{don}.
\item Ukawa presented the plans of the Center for Computational
  Sciences at Tsukuba for their new PACS-CS machine, which will
  consist of 2,560 2.8 GHz Intel Xeon processors connected by a
  three-dimensional Gbit-Ethernet-based hyper-crossbar network.  The
  peak performance of the machine is 14.3 TFlop/s, and it should be
  operational by July 2006.  For details, see Ref.~\cite{ukawa}.
\item KEK is currently collecting bids for a machine with at least 24
  TFlop/s peak performance, to be operational by March 2006.
\item A successor to QCDOC is under consideration, but concrete
  decisions have not yet been made.\\[-6mm]
\item IBM is working on a successor to BG/L called BlueGene/P.  This
  machine is supposed to break the PFlop/s barrier, but details on the
  architecture and the schedule have not yet been released.
\item Fujitsu has announced plans to build a machine with a peak
  performance of 3 PFlop/s by 2010/11.  This machine is supposed to
  use optical switching technology that has yet to be developed.
\item The Japanese technology ministry is discussing plans to develop
  a 10 PFlop/s machine by 2010/11, with a projected budget on the
  order of $10^{11}$ yen (roughly \$$10^9$ depending on the exchange
  rate).  A budgetary decision was supposed to be taken by the end of
  August but has been delayed because of the early elections in Japan.
\end{itemize}

Another interesting question is whether the Cell chip \cite{cell} can
be used for lattice QCD simulations.  This is a gaming chip made by
IBM for Sony and will power the Playstation 3.  It contains a new
PowerPC CPU and 8 FPUs, each of which has 256 kB of private memory.
There is also 512 kB of L2-cache on chip.  Each FPU can perform 4
single-precision FMAs per clock cycle.  At the nominal clock frequency
of 4 GHz, this translates into a peak performance of 256 GFlop/s
(single-precision) per chip.  Double-precision performance is
estimated to be about 10 times slower.  The memory interface (Rambus
XDR) has a bandwidth of 25.2 GB/s, and the I/O interface (Rambus
FlexIO) provides a total bandwidth of 76.8 GB/s (44.8 in, 32 out).
The big question for lattice QCD applications is whether the memory
system can keep the FPUs fed.  Since the chip has not yet been
released and details are hard to come by, only rough guesstimates are
possible.  Based on bandwidth estimates, a sustained single-node
performance on the order of 15-20\% might be achievable \cite{pab}.
Note, however, that the arithmetic is not IEEE-compliant (the FPUs
always round down).  This introduces a bias in the numerical Monte
Carlo simulations, which has to be eliminated by careful fix-up code.
In practice, this issue will set a limit on the amount of application
code that can take advantage of the high performance of the FPUs.

\section{Conclusions}
\label{sec:concl}

The machines that have been custom-designed for lattice QCD (apeNEXT
and QCDOC) are still the leading capability machines in our field.
They currently provide the best price-performance ratio (about \$1 per
sustained MFlop/s in the case of QCDOC) for large-scale lattice QCD
applications.  PC clusters are competitive as capacity machines and
might also be used as capability machines in the future as cluster
scalability improves.  If single precision is sufficient, their
cost-effectiveness is already comparable to that of the
custom-designed machines.  Commercial supercomputers are typically
much less cost effective.  It would not be a good idea for a lattice
group to purchase such a machine, but if computing time can be
obtained in a computing center it should of course be used.
BlueGene/L is an interesting alternative, as it is both a commercial
product and rather close to the QCDOC architecture.  To break even
with QCDOC, the purchase price for a rack of BG/L should not be more
than about \$1 million.  (Apparently, IBM has also been offering
computing time on BG/L at a price of \$6.7 million per rack and year
\cite{don}.)  Looking into the future, multi-core chips will become
prevalent very soon.  All three avenues (custom-designed machines, PC
clusters, and commercial supercomputers) should continue to be
explored, and all three of them are likely to remain important.

Designing or purchasing the best possible hardware is not sufficient.
Lean and efficient software is absolutely necessary to obtain high
performance.  The trend towards on-chip parallelization will pose new
challenges to the programmer, and the programming models will have to
change to make maximum use of multiple cores per chip.
High-performance code will most likely contain a combination of
pthreads, OpenMP, and MPI.  It will be very interesting to follow the
future developments in this area.

\section*{Acknowledgments}

The author would like to thank P.A. Boyle, N.H. Christ, M.A. Clark, Z.
Fodor, R. Edwards, S. Gottlieb, D. Holmgren, T. Lippert, D. Pleiter,
H. Simma, T.  Streuer, L. Tripiccione, and P. Vranas for helpful
correspondence and/or discussions.  This work is supported in part by
the Deutsche Forschungsgemeinschaft (DFG).


\begin{thebibliography}{99}
\bibitem{kennedy} see, e.g., T.A. Kennedy, Nucl. Phys. B (Proc.
  Suppl.) 140 (2005) 190
  [\href{http://arXiv.org/abs/hep-lat/0409167}{hep-lat/0409167}]
  and references therein
\bibitem{latfor} M. Hasenbusch et al., Nucl. Phys. B (Proc. Suppl.)
  129 (2004) 847 [\href{http://arXiv.org/abs/hep-lat/0309149}
  {hep-lat/0309149}] 
\bibitem{bagel} P.A. Boyle, in preparation
\bibitem{milc} \href{http://www.physics.utah.edu/~detar/milc}
  {http://www.physics.utah.edu/\~{}detar/milc}
\bibitem{chroma} \href{http://www.jlab.org/~edwards/chroma}
  {http://www.jlab.org/\~{}edwards/chroma} 
\bibitem{cps} \href{http://qcdoc.phys.columbia.edu/chulwoo_index.html}
  {http://qcdoc.phys.columbia.edu/chulwoo\_index.html}
\bibitem{fermiqcd} \href{http://www.fermiqcd.net}
  {http://www.fermiqcd.net} 
\bibitem{usqcd} \href{http://www.usqcd.org}{http://www.usqcd.org}
\bibitem{earthsim} S. Aoki et al., Annual Report of the Earth
  Simulator Center (2004),
  \href{http://www.es.jamstec.go.jp/esc/images/annualreport2003/pdf/project/chapter4/4-04ukawa.pdf}
  {http://www.es.jamstec.go.jp/esc/images/annualreport2003/pdf/project/chapter4/4-04ukawa.pdf}
\bibitem{altix} \href{http://www.uni-koeln.de/rrzk/kompass/104/k1045.html}
  {http://www.uni-koeln.de/rrzk/kompass/104/k1045.html}
\bibitem{bgl1} G. Bhanot, D. Chen, A. Gara, and P. Vranas,
  Nucl. Phys. B (Proc. Suppl.) 119 (2003) 114
  [\href{http://arXiv.org/abs/hep-lat/0212030}{hep-lat/0212030}] 
\bibitem{bgl2} G. Bhanot, D. Chen, A. Gara, J. Sexton, and P. Vranas,
  Nucl. Phys. B (Proc. Suppl.) 140 (2005) 823
  [\href{http://arXiv.org/abs/hep-lat/0409042}{hep-lat/0409042}] 
\bibitem{ibm} \href{http://www.research.ibm.com/journal/rd49-23.html}
  {IBM Journal of Research and Development 49 (2005)}
\bibitem{lippert} T. Lippert, Nucl. Phys. B (Proc. Suppl.) 129 (2004)
  88 [\href{http://arXiv.org/abs/hep-lat/0311011}{hep-lat/0311011}]
\bibitem{holmgren04} D.J. Holmgren, Nucl. Phys. B (Proc. Suppl.)
  140 (2005) 183 [\href{http://arxiv.org/abs/hep-lat/0410049}
  {hep-lat/0410049}]
\bibitem{holmgren05} D.J. Holmgren,
  \href{http://pos.sissa.it/archive/conferences/020/105/LAT2005_105.pdf}
  {PoS (LAT2005) 105} 
\bibitem{edwards} R.G. Edwards,
  \href{http://online.kitp.ucsb.edu/online/lattice_c05/edwards}
  {http://online.kitp.ucsb.edu/online/lattice\_c05/edwards}
\bibitem{holmgren} D.J. Holmgren,
  \href{http://lqcd.fnal.gov/allhands_holmgren.pdf}
  {http://lqcd.fnal.gov/allhands\_holmgren.pdf}
\bibitem{csikor} F. Csikor et al., Comput. Phys. Commun. 134 (2001) 139 
  [\href{http://arxiv.org/abs/hep-lat/9912059}{hep-lat/9912059}]
\bibitem{luescher} M. L\"uscher, Nucl. Phys. B (Proc. Suppl.) 106
  (2002) 21 
  [\href{http://arxiv.org/abs/hep-lat/0110007}{hep-lat/0110007}]
\bibitem{pochinsky} A. Pochinsky, Nucl. Phys. B (Proc. Suppl.)  140
  (2005) 859
\bibitem{mvia} \href{http://crd.lbl.gov/FTG/MVIA/mvia.shtml}
  {http://crd.lbl.gov/FTG/MVIA/mvia.shtml}
\bibitem{qmp} \href{http://www.lqcd.org/qmp}{http://www.lqcd.org/qmp}
\bibitem{steve} S. Gottlieb, private communication
\bibitem{don} D.J. Holmgren, private communication
\bibitem{grape} \href{http://www.astrogrape.org}{http://www.astrogrape.org}
\bibitem{apenext:lat04} F. Bodin et al., Nucl. Phys. B (Proc. Suppl.)
  140 (2005) 176 
\bibitem{apenext} F. Bodin et al., 
  \href{http://arxiv.org/abs/hep-lat/0306018}{hep-lat/0306018},
  \href{http://arxiv.org/abs/hep-lat/0309007}{hep-lat/0309007}
\bibitem{hubert} H. Simma, private communication
\bibitem{lele} L. Tripiccione, private communication
\bibitem{qcdsp} R.D. Mawhinney, Parallel Comput. 25 (1999) 1281
  [\href{http://arXiv.org/abs/hep-lat/0001033}{hep-lat/0001033}] 
\bibitem{qcdoc:lat04} P.A. Boyle et al., Nucl. Phys. B (Proc. Suppl.)
  140 (2005) 169, 829
\bibitem{qcdoc} P.A. Boyle, C. Jung, and T. Wettig,
  \href{http://arxiv.org/abs/hep-lat/0306023}{hep-lat/0306023};
  P.A. Boyle et al.,
  \href{http://www.research.ibm.com/journal/rd/492/boyle.pdf}
  {IBM Journal of Research and Development 49 (2005) 351}
\bibitem{nhc} N.H. Christ, private communication
\bibitem{callanan} O. Callanan, A. Nisbet, M. Peardon, and D. Gregg,
  \href{http://pos.sissa.it/archive/conferences/020/102/LAT2005_102.pdf}
  {PoS (LAT2005) 102}
\bibitem{apenet} R. Ammendola et al., 
  \href{http://pos.sissa.it/archive/conferences/020/100/LAT2005_100.pdf}
  {PoS (LAT2005) 100}
\bibitem{ukawa} A. Ukawa, 
  \href{http://pos.sissa.it/archive/conferences/020/111/LAT2005_111.pdf}
  {PoS (LAT2005) 111}
\bibitem{cell} There is very little official information available on
  the Cell chip.  The numbers quoted in the text were taken from
  \href{http://www.realworldtech.com/page.cfm?ArticleID=RWT021005084318}
  {http://www.realworldtech.com/page.cfm?ArticleID=RWT021005084318}
\bibitem{pab} P.A. Boyle, private communication
\end{thebibliography}
\end{document}